\documentclass[aps,prapplied,twocolumn,superscriptaddress,longbibliography]{revtex4-2}

\usepackage{babel}
\usepackage{graphicx}
\usepackage{bm}
\usepackage{color}
\usepackage{epstopdf}
\usepackage{amsmath} 
\usepackage{url}
\definecolor{darkGreen}{RGB}{0,110,0}
\definecolor{darkBlue}{RGB}{0,0,130}
\usepackage[colorlinks=true, urlcolor=darkBlue,citecolor=darkGreen,linkcolor=darkBlue]{hyperref}
\usepackage{amssymb}
\usepackage{comment}
\usepackage{physics}

\begin{document}


\title{Memory-Augmented Hybrid Quantum Reservoir Computing}



\author{J. Settino}
\affiliation{Dipartimento di Fisica, Università della Calabria, Via P. Bucci Arcavacata di Rende (CS), Italy}
\affiliation{INFN, Gruppo collegato di Cosenza, Italy}
\affiliation{ICAR-CNR, Consiglio Nazionale delle Ricerche, Italy}
\email[]{jacopo.settino@unical.it}

\author{L. Salatino}
\affiliation{ICAR-CNR, Consiglio Nazionale delle Ricerche, Italy}

\author{L. Mariani}
\affiliation{ICAR-CNR, Consiglio Nazionale delle Ricerche, Italy}

\author{F. D'Amore}
\affiliation{ICAR-CNR, Consiglio Nazionale delle Ricerche, Italy}

\author{M. Channab}
\affiliation{NTT Data Via Calindri, 4 Milano, Italy}
\author{L. Bozzolo}
\affiliation{NTT Data Via Calindri, 4 Milano, Italy}
\author{S. Vallisa}
\affiliation{Fondazione ISI via Chisola 5 Torino}
\author{P. Barillà}
\affiliation{NTT Data Via Calindri, 4 Milano, Italy}
\author{A. Policicchio}
\affiliation{NTT Data Via Calindri, 4 Milano, Italy}

\author{N. Lo Gullo}
\affiliation{Dipartimento di Fisica, Università della Calabria, Via P. Bucci Arcavacata di Rende (CS), Italy}
\affiliation{INFN, Gruppo collegato di Cosenza, Italy}
\affiliation{ICAR-CNR, Consiglio Nazionale delle Ricerche, Italy}

\author{A. Giordano}
\affiliation{ICAR-CNR, Consiglio Nazionale delle Ricerche, Italy}

\author{C. Mastroianni}
\affiliation{ICAR-CNR, Consiglio Nazionale delle Ricerche, Italy}

\author{F. Plastina}
\affiliation{Dipartimento di Fisica, Università della Calabria, Via P. Bucci Arcavacata di Rende (CS), Italy}
\affiliation{INFN, Gruppo collegato di Cosenza, Italy}

\date{\today}


\date{\today}
 
\begin{abstract}

Reservoir computing is an effective method for predicting chaotic systems by using a high-dimensional dynamic reservoir with fixed internal weights, while keeping the learning phase linear, which simplifies training and reduces computational complexity compared to fully trained recurrent neural networks.
Quantum reservoir computing uses the exponential growth of Hilbert spaces in quantum systems, allowing for greater information processing, memory capacity, and computational power. 
We present a hybrid neuromorphic quantum-classical approach that implements memory through classical post-processing of quantum measurements, thus avoiding the need for multiple coherent input injections (as in the original proposal). We tested our model on two physical platforms: a fully connected Ising model and a Rydberg atom array, and evaluated it on various benchmark tasks, including the chaotic Mackey-Glass time series prediction, where it demonstrates significantly enhanced predictive capabilities and achieves a substantially longer prediction time, outperforming previously reported approaches.
\end{abstract}


\maketitle


\section{Introduction}

Reservoir Computing (RC) is a computational framework that has gained significant attention for efficiently processing time-dependent data. Originally inspired by recurrent neural networks (RNNs), RC exploits the dynamics of a fixed, high-dimensional system, known as the reservoir, to transform input signals into a rich, dynamic representation \cite{Maass2002,Jaeger2004}. This transformation allows for the linear extraction of complex features, making RC particularly suitable for tasks such as time series prediction, pattern recognition, and control systems. The key advantage of RC lies in its simplicity and efficiency. Unlike traditional RNNs, where the entire network is trained, RC requires training only on the output layer, while the reservoir remains untrained\cite{Maass2002, Jaeger2004,Gauthier2021, Milano2023,Milano2023a,Ma2023,Xiao2021}. 
RC systems can be implemented in the quantum domain, exploiting quantum mechanics properties to further enhance computational power and memory capacity. Indeed, quantum computing offers new information processing paradigms by exploiting the superposition principle, entanglement, and quantum interference, allowing for exponential increases in computational power and memory space. These features enable quantum systems to tackle problems that are computationally infeasible for classical ones \cite{Shor1994, Grover1996, Nielsen2010, Montanaro2016, Preskill2018,Mastroianni2022, Mastroianni2023,Mastroianni2023Numta,Mastroianni2024,Consiglio2024,DeLorenzis2024,Boneberg2023}. The first proposal of Quantum Reservoir Computing (QRC) by Fujii and Nakajima \cite{Fujii2017} utilizes the space of quantum density matrices to store and process information, by injecting the input data into the state of one qubit of a fully connected qubit network.

Building on that foundational work, researchers have investigated a variety of platforms, each offering specific advantages for Quantum Reservoir Computing (QRC) \cite{Mujal2021,Ghosh2021,Abbas2024,Zhu2024}. Among these, spin chains have been extensively studied across a wide range of tasks, from time series prediction to Extreme Learning Machines \cite{Mujal2022,Sannia2024,Mifune2024,Xia2022,Xia2023,Martínez2021,Martínez2023,Gotting2023,Kobayashi2024,Innocenti2023,LoMonaco2024,Vetrano2024,Palacios2024,Kobayashi2024c,Monzani2024,Ivaki2024,Ahmed2024a}.
Additionally, fermionic and bosonic systems have been proposed as alternative physical implementations of reservoir \cite{Ghosh2019,HoanTran2023,Llodra2023}. Other explored set-ups include quantum oscillators \cite{Govia2021,Dudas2023}, photonic systems for both QRC \cite{Giorgi2024}, and Quantum Extreme Learning Machines \cite{Suprano2024}, and Rydberg atoms, which offer strong dipole-dipole interactions as a key feature \cite{Bravo2022,Kornjača2024}, with platforms such as Bloqade by QuEra enabling the simulation of many-body quantum systems. Furthermore, gate-based platforms have recently been explored for implementing QRC algorithms, due to their universality and high controllability \cite{Domingo2023,Domingo2023a,Yasuda2023,Wudarski2023,Fry2023,Kubota2023,Fuchs2024,Kobayashi2024a}.

Despite their theoretical promise, protocols based on Fujii's approach require multiple copies of the quantum system, rendering them experimentally impractical. To address this challenge, hybrid approaches have been proposed \cite{Wudarski2023,HoanTran2023,Kobayashi2024a,Ahmed2024}, drawing inspiration from classical reservoir computing \cite{Tanaka2019,Nakajima2020,Cucchi2022,Platt2022}. In these hybrid models, quantum evolution is combined with classical post-processing, providing a more feasible solution for practical implementation. However, these hybrid systems often suffer from a significant drawback: the loss of memory of the previous steps. Feedback mechanisms, wherein the result of the previous step is reintroduced into the system at the subsequent step, have been proposed to address this issue \cite{Wudarski2023,Kobayashi2024,Ahmed2024,Nokkala2024}. 

\begin{figure}[h!btp]
    \centering
    \includegraphics[width=0.95\linewidth]{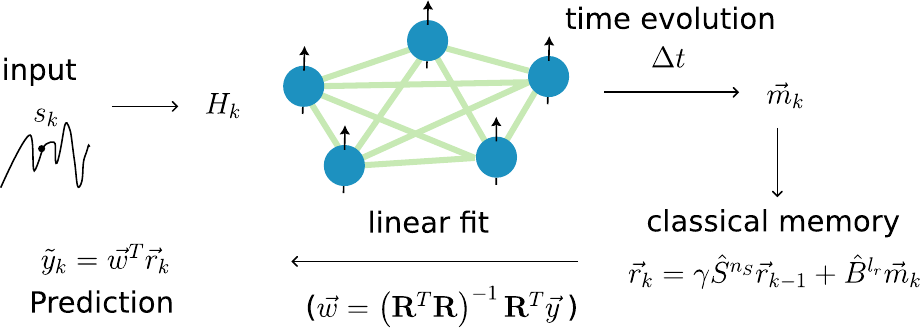}
    \caption{\small{Schematic representation of the proposed hybrid quantum-classical reservoir computing (QRC) method. At each step $k$, the input data $\{s_k\}$ is encoded into the quantum reservoir by modulating a parameter of the Hamiltonian $H_k$, which generates the dynamics of a quantum system (e.g., a transverse-field Ising model or Rydberg atom array). The system evolves for a time $\Delta t$, giving rise to the quantum state $|\psi_k(\Delta t)\rangle$, depending on the input. After the time evolution, the output is obtained by measuring a set of observables $\vec{m}_k$, which carry information about the input data. These measurements are further processed classically to construct a reservoir state $\vec{r}_k$ that retains the memory of previous inputs. A linear fit is then applied to the reservoir state to produce the predicted output, thereby allowing the system to solve complex tasks such as time series prediction with enhanced memory capacity.}}
    \label{fig:scheme}
\end{figure}

In this work, we introduce a novel quantum-classical hybrid approach that integrates a simple classical post-processing component to boost memory capacity by exploiting classical memory resources (see Fig.~\ref{fig:scheme}), bridging between QRC and Extreme Learning Machine. We assess the performance of our algorithm across several benchmark tasks, including linear memory, NARMA, and, finally, time series prediction, using the Mackey-Glass series as an example. Our approach is tested on two different physical systems: a fully connected Ising model in a transverse field and a system utilizing Rydberg atoms. For both of the cases, we find a substantial enhancement of predictive capabilities.

The paper is structured as follows. In Sec.~\ref{sec:Method}, we introduce the hybrid quantum-classical reservoir computing model and detail the methods used to enhance memory capacity through classical post-processing. Sec.~\ref{sec:Results} presents the implementation of our model on two distinct physical systems: a fully connected Ising model and a Rydberg atoms platform. Finally, in Sec.~\ref{sec:Conclusions}, we summarize our findings and discuss future directions for the development of quantum reservoir computing. Further details the Rydberg‐atom Hamiltonian are given in Appendix A, while Appendix B presents the system-size scaling study,  Appendix C analyzes measurement-noise effects, and Appendix D benchmarks our QRC against a classical ESN.

\section{Method}\label{sec:Method}

We introduce a model of quantum reservoir computing where the input is encoded into the system by varying a Hamiltonian parameter. Our model uses a Hamiltonian similar to that proposed by Fujii and Nakajima \cite{Fujii2017}; namely, a transverse-field Ising model with an additional input-dependent magnetic field aligned with the coupling terms:

\begin{equation}
\hat{H_k} = \sum_{i<j} J_{ij} \hat{\sigma}^x_i \hat{\sigma}^x_j + h \sum_i \hat{\sigma}^z_i + h_k \sum_i \hat{\sigma}^x_i,
\end{equation}

where  \( J_{ij} \) represents the coupling constants and \( h \) is the static magnetic field. The main difference between our model hamiltonian and previous approaches \cite{Fujii2017,Sannia2024} lies in the inclusion of an input-dependent magnetic field term \(h_k\) that directly encodes the \(k\)-th input value \(h_k=C_s s_k\).
 For each input $s_k$, a target function  $y_k(\{s_{k'}\}_{k'\leq k})$ is identified, which generically depends on the previous values of the input sequence.


 For each input data, a pure state $  \ket{\psi_0} $ is evolved for a time $\Delta t $ according to the  Hamiltonian $\hat{H}_k$,  $\ket{\psi_0^k(\Delta t)}=e^{-i\hat{H}_k \Delta t} \ket{\psi_0}$. After this evolution, the expectation values of a set of observables,  $\vec{m_k}$, are measured. 


Since the evolution is driven only by the current input \(s_k\), we aim to construct a reservoir state that retains memory of previous states. Inspired by the structure of classical RNNs \cite{Cucchi2022,Nakajima2020,Platt2022,Tanaka2019}, we define the ``reservoir state (vector)" by incorporating two additional operators (\(\hat{S}^{n_S}\) and \(\hat{B}^{l_r}\)) into the classical framework:

 \begin{equation}
     \vec{r}_k=\gamma \hat S^{n_S} \vec{r}_{k-1} + \hat B^{l_r} \vec{m}_k.
 \end{equation}
$\hat S^{n_S}$ and $\hat B^{l_r}$ act on the reservoir state at the previous step and on the measurement vector, respectively, with $\gamma$ being a real-valued parameter between $0$ and $1$ that weighs the previous input. $\hat S^{n_S}$ acts on a generic vector by shifting the vector elements by $n_S$ steps forward, if positive, or backward, if negative, (${\left[\hat S^{n_S} \right]}_{i j}=\delta_{\left(i+n_S\right) \bmod l_r, j}$). It coincides with a cyclic permutation of the reservoir vector and can be represented by an orthogonal matrix. Its role is to enhance the distinguishability of the inputs at different times by avoiding the same observable being summed up at different times, thus boosting classical memory. The operator $\hat B^{l_r}$ acts on the measurement vector by lengthening it, interleaving zeros between the elements such that its final length is $l_r$, which coincides with the length of $\vec{r_k}$. 
As an example, given a vector \( \{v_1, v_2, v_3,v_4\} \), the action of  the operator \( \hat{S^1} \) coincides with \( \hat{S}^1\{v_1, v_2, v_3,v_4\} = \{v_2, v_3,v_4, v_1\} \), while applying \( \hat{B}^8 \) introduces four  zeros, resulting in \( \hat{B}^8\{v_1, v_2, v_3,v_4\} = \{v_1, 0,  v_2, 0, v_3, 0, v_4, 0 \} \).

In the training phase, after the reservoir is built with the measurements coming from $N_{tr}$ values of the input sequence, a linear fit is applied. Specifically, the predicted value at step $k$ is given by $\tilde{y}_k=\vec{w}^T \vec{r}_k$, with the weight vector being given by:
\[
\vec{w} = \left(\mathbf{R}^T \mathbf{R} \right)^{-1} \mathbf{R}^T \vec{y},
\]
where \( \mathbf{R} \) is the matrix of reservoir vectors, \( \mathbf{I} \) is the identity matrix, and \( \vec{y} \) is the vector of target values on the training dataset. 

To simulate the Reservoir Computing algorithm, we perform exact diagonalization of $N$=$5$ spins,  with $J_{i j}$ randomly generated from a uniform distribution in \([-J, J]\), where we take $J$ as our energy and inverse time unit. Other parameters are $h$=$0.2 J$, $C_s=0.2 J$, $\gamma=0.95$ for the memory tasks and $\gamma=0.6$ for the prediction task. For the memory task, $1000$ points are used for training and an equal number for testing, discarding the first $100$ steps as burn-in phase; for the prediction tasks, $1500$ points are used for training and up to $1000$ for the testing, discarding the first $200$ steps as burn-in phase. The initial state is chosen as the pure state  $\ket{\psi_{0}} = \ket{\uparrow\uparrow\ldots\uparrow}$  (all spins are in the eigenstate of $\sigma_z$ corresponding to the positive eigenvalue, i.e., the $\lvert \uparrow \rangle$ state). When not specified, $n_S=1$ and the lengthening operator is not used ($\hat B^{l_r}=I$). A table with names for the sets of observables is provided below in Tab.~\ref{table:observables}.

\begin{table}[h!]
\centering
\begin{tabular}{|c|c|c|}
\hline
Set Name & Observables & Cardinality\\ \hline
$\vec{m}^{X 1}$& \(\left\{\expval{\sigma^x_1}\right\}\)& 1 \\ \hline
$\vec{m}^{1}$& $ \left\{ \expval{\sigma^x_1}, \expval{\sigma^y_1},\expval{\sigma^z_1} \right\} $& 3\\ \hline
$\vec{m}^{N}$& $\bigcup_{j=1}^{N} \left\{ \expval{\sigma^x_j}, \expval{\sigma^y_j}, \expval{\sigma^z_j} \right\}$& 15\\ \hline
$\vec{m}^{XX}$ & $ \left\{ \expval{\sigma_{i}^{x}} \right\}_{i=1, \ldots, N} \cup \left\{ \expval{ \sigma_{i}^{x} \sigma_{j}^{x}}\right\}_{i<j} $& 15\\ \hline
$\vec{m}^{1-N}$& $  \left\{ \expval{\sigma^x_1}, \expval{\sigma^y_1},\expval{\sigma^z_1} \right\}  \cup \left\{ \expval{ \sigma_{1}^{\alpha} \sigma_{j}^{\alpha}}\right\}_{1<j}^{\alpha \in\{x, y, z\}} $& 15\\ \hline
$\vec{m}^{A}$& $\left\{\left\langle\sigma_{i}^{\alpha}\right\rangle\right\}_{i=1, \ldots, N}^{\alpha \in\{x, y, z\}} \cup\left\{\left\langle\sigma_{i}^{\alpha} \sigma_{j}^{\alpha}\right\rangle\right\}_{i<j}^{\alpha \in x, y, z}
$& 45\\ \hline

\end{tabular}
\caption{Sets of observables used in the analysis. In the ``Observables'' column, the superscript refers to the Pauli Matrix index, while the subscript refers to the position index of the individual spin (or qubit) in the system.  }
\label{table:observables}
\end{table}

\section{Results}\label{sec:Results}

In this section, we evaluate the performance of our hybrid quantum-classical reservoir computing model. We first test the short-term memory (STM), followed by the Nonlinear Auto-Regressive Moving Average (NARMA) task. We then assess the model's predictive capabilities on the chaotic Mackey-Glass time series. Lastly, we validate the generality of our approach by simulating the model on a Rydberg atom platform.

\subsection{STM task}

To begin the analysis of the performance of our Reservoir Computing (RC) algorithm, we examine the simplest task: assessing the system's short-term memory (STM) capabilities, which measures the reservoir's ability to store and retrieve past input sequences. In this task, the input sequence  \(\{s_k\}\) is randomly generated from a uniform distribution over the interval \([-1, 1]\). The goal is to obtain the input value from \(\tau\) time steps earlier: \(y_k = s_{k-\tau}\).

Performance is measured by the Squared Pearson Correlation Coefficient (SPCC) for each delay \(\tau\), defined as:

\begin{equation}
\text{C}(\tau) = \frac{\text{cov}(y_k, \hat{y}_k)^2}{\text{var}(y_k) \text{var}(\hat{y}_k)},
\end{equation}
where \(\hat{y}_k\) is the predicted output, ``\text{cov}'' denotes covariance, and ``\text{var}'' denotes variance. This metric quantifies the reservoir’s ability to recall inputs delayed by $\tau$ steps.

\begin{figure} 
    
    \centering

        \includegraphics[width=\linewidth]{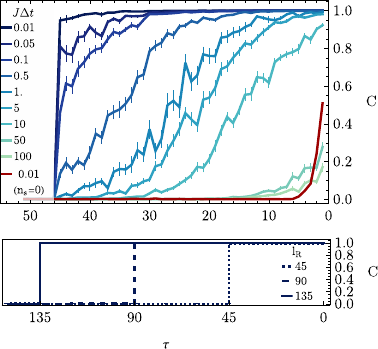}

    \caption{       \label{fig:STM1} \small{Top) SPCC for the STM task as a function of $\tau$ for several $\Delta t$ values of the Hamiltonian evolution. The set of observables $\vec{m}^{A}$ is used. The cyclic permutation operator is employed in all cases (with $n_s=1$), except for the red line, for which $n_s=0$; the lengthening operator is not used ($\hat B^{l_r}=I$).  Bottom) SPCC as a function of $\tau$ for several reservoir lengths $l_r$ determined by the $\hat B^{l_r}$ operator when only one observable is used ($\vec{m}^{X1}$). The cyclic permutation operator is implemented with $n_s=1$. Each point in the plot shows the mean taken over $30$ different realizations of the Hamiltonian, while the bar is the standard error on the mean.}
}
\end{figure}

In Fig.~\ref{fig:STM1} (top), we present the SPCC as a function of the delay \(\tau\) for different \(\Delta t\) values, for the case with the cyclic permutation operator ($n_s$=$1$), when using the set of observables $\vec{m}^{A}$. One can notice that the SPCC increases with decreasing \(\Delta t\), as the simplicity of the task does not require the nonlinearity introduced by longer time evolutions. For the smallest \(\Delta t\), the STM remains nearly perfect up to $\tau$=$45$, corresponding to the reservoir dimension.

 As a comparison, we also display the case without the cyclic permutation operator (\(n_s=0\)), red line in Fig.~\ref{fig:STM1}, where  $\Delta t$ has been set to its best value ($J \Delta t=0.01$). It is evident that the cyclic permutation operator is crucial for retaining significant memory: the red line is almost zero for $\tau>2$.

In Fig.~\ref{fig:STM1} (bottom), we observe that when using the observable set $\vec{m}^{X 1}$, which contains only one observable on a single spin, and artificially increasing the reservoir dimension through the lengthening operator \(B^{l_r}\), it is possible to extend the STM capacity arbitrarily. This occurs because the classical memory updating function, combined with the permutation operator, allows for the storage of \(l_r\) expectation values of \(\expval{\sigma^1_x}\), which contain the injected inputs in a way that the linear fit can distinguish them. It is worth noting that the optimal performance ($C=1$ for $\tau \le l_R$)\cite{Farkas2016} is already obtained with a single observable; thus, it is not necessary to measure more.


\subsection{NARMA task}

To further check the performance of our RC algorithm, we test it for the Nonlinear Auto-Regressive Moving Average (NARMA) task, which is a benchmark usually employed to evaluate the non-linear memory and computational capabilities of RC systems. It involves generating a sequence where the next value depends non-linearly on both the previous ones and a random input sequence. The NARMA task with order $n$ is defined as:

\begin{multline}
y_k = 0.3 y_{k-1} + 0.05 y_{k-1} \left( \sum_{i=0}^{n-1} y_{k-i-1} \right) \\ + 1.5 s_{k-n} s_{k-1} + 0.01,
\end{multline}

where \( s_k \) is a random input uniformly distributed in the interval \([0, 0.1]\), and \( y_k \) is the target output. The goal is to train the reservoir system to reproduce \( y_k \) given the input sequence \( s_k \). 

The performance on the NARMA task is measured by the accuracy with which the reservoir reproduces the target output sequence; to quantify it,  we use again the SPCC. 

\begin{figure}
    \centering
        \includegraphics[width=\linewidth]{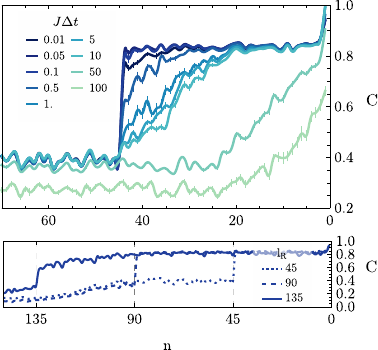}
    \caption{       \label{fig:Narma} \small{Top) SPCC for the NARMA task, as a function of the NARMA order $n$ for several $\Delta t$ values of the Hamiltonian evolution. The set of observables $\vec{m}^{A}$ is employed and the lengthening operator is not used ($\hat B^{l_r}=I$). Bottom) C as a function of $n$ for several reservoir lengths $l_r$, determined by the $\hat B^{l_r}$ operator when only one observable is used ($\vec{m}^{X1}$). In both panels, the cyclic permutation operator is implemented with $n_s=1$.}
    The mean and the standard error on the mean are shown for each point with $30$ different realizations of the Hamiltonian.}
\end{figure}

Fig.~\ref{fig:Narma} (top) shows that, in contrast to the linear memory task, the performance on the NARMA task does not reach its peak for the smallest \(\Delta t\), but for $J \Delta t\approx 0.05$. 

When compared to the STM task, the NARMA task is more complex because it includes products of the input sequence's values, at different time steps. As a consequence, the quantum system needs to evolve for a longer time to chase a more complex function of the input.

In the top panel, we observe that once the NARMA order \(n\) exceeds the reservoir dimension \(l_R = 45\), the SPCC falls off sharply, revealing the finite memory limit of the network.
Also in this case, the reservoir memory can be significantly enhanced by increasing the reservoir dimension using the lengthening operator \(B^{n_B}\), as illustrated in Fig.~\ref{fig:Narma} (bottom). 

\begin{figure}[h!btp]
    \centering
    \includegraphics[width=0.9\linewidth]{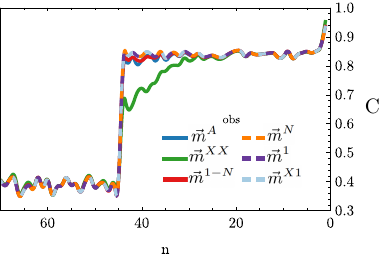}
    \caption{\small{SPCC for the NARMA task as a function of the NARMA parameter $n$ for several sets of observables (see Tab.~\ref{table:observables}). Reservoir length is fixed as ${l_r=45}$. The cyclic permutation operator is implemented with ${n_s=1}$.
    }
    }
    \label{fig:N2}
\end{figure}

In Fig.~\ref{fig:N2}, we examine the SPCC for the NARMA task using different sets of observables, while keeping the reservoir dimension fixed at ${l_r=45}$. The best results, as in the STM case, are achieved with a single observable (\(\vec{m}^{X 1}\)), which is experimentally more friendly, as it simplifies the implementation and reduces the complexity of the measurement process.
Indeed, it is seen that increasing the number of observables causes a small decrease in performance. The reason lies in the fact that, with a fixed reservoir dimension, a larger number of observables leaves less room for classical memory.
   
\subsection{Prediction task}

Next, we move on to a prediction task, which is one of the primary applications of reservoir computing. This shift in focus allows us to evaluate the system's capability to forecast future values based on its internal states, which depend on the previous input data. 
We tested our model with the prediction of a chaotic time series, i. e. the Mackey-Glass (MG) time series. The MG series \( s_k \) is defined by the following delay differential equation:

\begin{equation}
\frac{ds(t)}{dt} = \beta \frac{s(t - \tau)}{1 + s(t - \tau)^n} - \Gamma s(t)
\end{equation}

The differential equation is numerically solved and the solution is sampled with a time step \( \delta t \), to define the discrete sequence. Parameters are fixed as in Ref.~\cite{Sannia2024}:  
$\beta$=$0.2$, $ \Gamma$=$0.1$,  delay $\tau$=$17$ (chaotic behavior emerges for values greater than $17$),  $\delta t$=$3.0$, non-linearity parameter $n$=$10$. In the training phase, we implement a  one-step forward prediction task, with $y_k$=$s_{k+1}$. In the test phase, starting from the last training time, a longer-time prediction is obtained by multiple one-step predictions.  

To evaluate the performance of our reservoir model, we use the Valid Prediction Time (VPT), defined as follows \cite{Wudarski2023}. 
The VPT is the maximum time \( T \) for which the predicted trajectory \( \hat{y}(t) \) remains within an acceptable error margin from the true trajectory \( y(t) =s(t)\). Mathematically, this can be expressed as:

\begin{equation}
VPT = \max \left\{ T : \forall t \leq T, \left(\frac{\tilde{y}(t)-y(t)}{\sigma}\right)^2 < \varepsilon^2, \right\}
\end{equation}
where \( \epsilon=0.3 \) is a predefined error tolerance and $\sigma$ is the standard deviation of the time series. We express VPT in integer units of the discretized time.   

\begin{figure}[h!btp]
    \centering
    \includegraphics[width=0.9\linewidth]{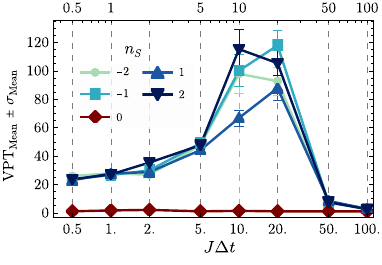}
    \caption{
    \small{VPT for the MG series prediction as a function of \(\Delta t\) in the Hamiltonian evolution for different shift numbers \(n_s\). For each point, the mean value and standard deviation of the mean are shown, based on $30$ different realizations of the Hamiltonian. The set of observables $\vec{m}^{A}$ is employed and the lengthening operator is not used ($\hat B^{l_r}=I$).
}}
    \label{fig:MG1}
\end{figure}

From Fig.~\ref{fig:MG1}, it is evident that the cyclic permutation function is crucial, as the prediction capability is negligible without it, and there are slight performance differences between various permutation functions, i.e.,  $n_s$=$\pm 1,\pm 2$. The best evolution time for the spin network is \(J \Delta t \approx 10\), significantly longer than the time required for memory tasks. This indicates that more complex tasks necessitate greater non-linear processing and signal distribution across the network nodes. The extended evolution time allows the system to better integrate and process the temporal dependencies inherent in the MG equation, thereby enhancing prediction accuracy.


\begin{figure}[h!btp]
    \centering
    \includegraphics[width=0.9\linewidth]{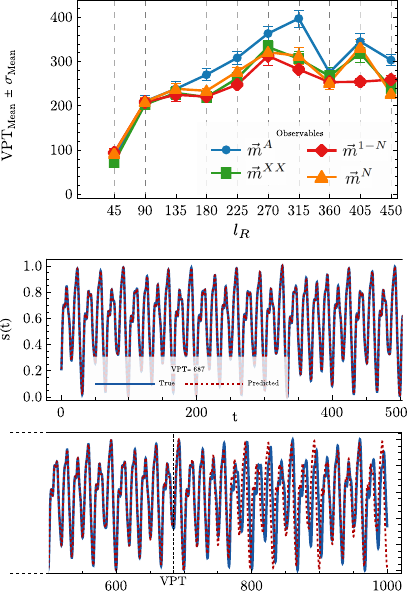}
\caption{Top) \small{VPT for the MG series prediction as a function of the reservoir dimension \(\l_r\) for different sets of observables. For each point, the mean value and standard deviation of the mean are shown, based on $30$ different realizations of the Hamiltonian.
Bottom) Example of the predicted time series for the best reservoir dimension ($l_r=315$) as a function of the integer discretized time.  }
\label{fig:MG2}}
\end{figure}

Setting \(\Delta t\) at its best value, Fig.~\ref{fig:MG2} (top) shows the VPT as a function of the reservoir dimension, adjustable via the operator $\hat B^{l_r}$, for different sets of observables. Firstly, a general improvement in VPT is observed with increasing the reservoir dimension for all the sets of observables, up to a saturation point around $270$.
Good performance is achieved with commutating observables ($\vec{m}^{XX}$) or single-particle observables ($\vec{m}^{N}$), while non-commutating and double-spin observables do not offer benefits ($\vec{m}^{1-N}$). Conversely, using the biggest considered set of observables ($\vec{m}^{A}$) provides a slight improvement in VPT. In Fig.~\ref{fig:MG2} (bottom), we present the true and predicted MG series for a specific case, with the set of observables $\vec{m}^{A}$. It is seen that in a specific realization of the Hamiltonian, the system can predict up to approximately 700 steps. Overall, the prediction capability of our hybrid algorithm is significantly higher than recent literature results \cite{Sannia2024,Kobayashi2024a}.

\subsection{Simulations with Rydberg atoms}

To further validate the generality of our method, we tested our algorithm on a different Hamiltonian, which is tailored to simulate the interactions in Rydberg atom arrays \cite{Bravo2022,Henriet2020}. 
 This Hamiltonian provides a robust platform for examining the performance of quantum reservoir computing in a different physical setting, ensuring that our method is not limited to a specific type of quantum system.

The Hamiltonian in terms of spin operators is given by:

\begin{equation}\begin{split}
    {H^R_k} & = \sum_j\Omega_j \left[ \sigma_j^x cos({\varphi_j}_k)- \sigma_j^y sin({\varphi_j}_k)\right] 
    \\ & - \sum_j \frac{{\Delta_j}_k}{2} - \sum_j \frac{{\Delta_j}_k}{2} \sigma_j^z 
    \\ & + \sum_{j < k} V_{jk} \frac{(1+\sigma_j^z)}{2} \frac{(1+\sigma_k^z)}{2}. 
\end{split}\end{equation}

where \(\Omega_j(t)\) is the Rabi frequency, \(\varphi_j(t)\) is the laser phase, \(\Delta_j(t)\) is the detuning and \(V_{jk}\) is the interaction term between atoms \(j\) and \(k\).  The term \( V_{jk} = \frac{C_6}{|\mathbf{x}_j - \mathbf{x}_k|^6} \) describes the van der Waals interaction, between atoms \( j \) and \( k \). Here, \( \mathbf{x}_j \) denotes the position of the \( j \)-th atom, and the denominator \( |\mathbf{x}_j - \mathbf{x}_k| \) represents the distance between atoms \( j \) and \( k \). The atoms are placed in a 1D-lattice, equally spaced with a minimum distance set to \(7 \mu\text{m}\) \cite{Bravo2022}. The coefficient \( C_6 \) is a constant that characterizes the strength of the interaction and depends on the specific Rydberg states being used. For our simulation, the default value of \( C_6 \) is set to \( 862690 \times 2\pi \ \text{MHz} \ \mu\text{m}^6 \). This value corresponds to the interaction strength for the Rydberg state \(|r \rangle = | 70S_{1/2} \rangle\) of \(^{87}\text{Rb}\) atoms.
According to Ref.~\cite{Bravo2022}, we further set \(\Omega = 2\pi \times 4.2\ \text{MHz}\).

\begin{figure}[h!btp]
    \centering
    \includegraphics[width=0.9\linewidth]{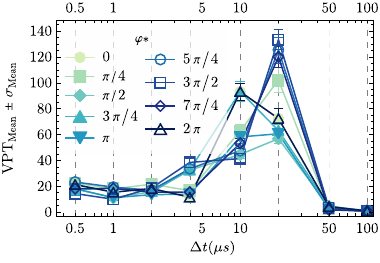}
    \caption{\small{VPT for the MG series prediction as a function of $\Delta t $ of the Rydberg-atoms Hamiltonian for different values of the phases $\varphi*$. The signal is encoded homogenously in the value of the detuning parameter \({\Delta_j}_k=C_s s_k\). For each point, the mean value and standard deviation of the mean are shown, based on $30$ different realizations of the Hamiltonian.}}
    \label{fig:MGR1}
\end{figure}

\begin{figure}[h!btp]
    \centering
    \includegraphics[width=0.9\linewidth]{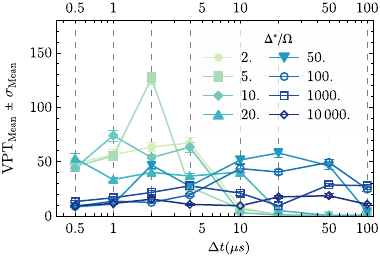}
    \caption{\small{VPT for the MG series prediction as a function of $\Delta t $ of the Rydberg-atoms Hamiltonian for different values of the detuning parameter $\Delta*$. The signal is encoded homogenously in the value of the phase \({\varphi_j}_k=\frac{\pi}{2} s_k\). For each point, the mean value and standard deviation of the mean are shown, based on $30$ different realizations of the Hamiltonian.}}
    \label{fig:MGR2}
\end{figure}

The initial state is chosen as the pure state  $\ket{\psi_{0}}$, and both the phases $\varphi_j$ and the parameters $\Delta_j$ are, in principle, capable of encoding the input information. We assessed the performance of our reservoir computing algorithm using both of these encoding methods and the set of observables \(\vec{m}^{A}\).
The information about the \( k \)-th value of the input sequence \(\{s_k\}\) is encoded homogeneously, first in the value of the detuning parameter \({\Delta_j}_k=C_s s_k\), by fixing the value of the phases to $\varphi^*$ (Fig.~\ref{fig:MGR1}), and subsequently in the phase \({\varphi_j}_k=\frac{\pi}{2} s_k\), by fixing the value of  ${\Delta_j}_k$ to ${\Delta^*}$ (Fig.~\ref{fig:MGR2}). 
In the first case, it is obtained that the maximum value of the \(VPT_{Mean}\) occurs when the influence of the spin component along the \(x\)-axis is nullified, and the component along the \(y\)-axis is maximized, thereby optimizing the interaction of the system with the external electromagnetic field. Suboptimal values are obtained for phases near $3 \pi /4$. 
In the second case, the dependence on the hyper-parameter $\Delta^*$ is more critical, the best performance is peaked at $\Delta^*= 5  \Omega$ for a specific evolution time $\Delta t= 2$.

The results obtained from this validation indicate that our approach retains its efficacy across different quantum systems, including those involving Rydberg atoms. The adopted Hamiltonian captures the essential features of Rydberg atom dynamics, such as the strong dipole-dipole interactions and long coherence times, which are critical for quantum information processing. Our findings suggest that the combination of classical operators for memory enhancement with the inherent quantum dynamics of Rydberg atoms can significantly improve the performance of quantum reservoir computing systems.

These results highlight the versatility of our method, demonstrating its potential applicability to a wide range of quantum systems and confirming its robustness in handling different quantum Hamiltonians.

\section{Conclusions}\label{sec:Conclusions}

In this work, we have developed a hybrid quantum-classical reservoir computing model that effectively combines the quantum dynamics of a system with classical post-processing techniques to enhance memory capacity. Our approach was designed to mitigate the challenges associated with purely quantum reservoir computing, such as requiring multiple copies of the quantum system. 

We demonstrated the efficiency of our model by applying it to two distinct physical systems: a fully connected Ising model in a transverse field and a Rydberg atom array. Both platforms showcased the versatility of our method, as they were able to successfully process information in a variety of benchmark tasks, including linear memory, NARMA, and the prediction of the chaotic Mackey-Glass time series. Our model outperformed several standard approaches reported in the literature~\cite{Sannia2024,Kobayashi2024a} across all tested benchmarks, demonstrating its advanced predictive capabilities and enhanced memory. Moreover, the performance gains achieved through system-size scaling (Appendix~\ref{app:scaling}) and the comparison with ESNs (Appendix~\ref{app:esn}) underscore the potential of deploying our approach on real quantum hardware for tackling real-world problems.
  
The key innovation of our approach lies in the introduction of two operators in the classical post-processing phase, allowing us to extend arbitrarily the classical system’s memory while maintaining the benefits of quantum information processing.
The proposed method allows a continuous transition from an Extreme Learning Machine when $\gamma = 0$ to QRC with a tunable memory via the parameter $\gamma$. This tunability enables optimizing the system's memory capacity according to the specific requirements of the target applications.
This hybrid approach provides a promising pathway for future reservoir computing applications, particularly in cases where quantum resources are limited or expensive to replicate experimentally.



\section{Acknowledgements}
We acknowledge support from PNRR MUR project PE0000023-NQSTI through the secondary projects “QuCADD” and “ThAnQ”. JS acknowledges the contribution from PRIN (Progetti di Rilevante Interesse Nazionale) TURBIMECS - Turbulence in Mediterranean cyclonic events, grant n. 2022S3RSCT CUP H53D23001630006, CUP Master B53D23007500006, CM from ICSC – Italian Research Center on High Performance Computing, Big Data and Quantum Computing, funded by European Union – NextGenerationEU, PUN: B93C22000620006.

\vspace{20cm}

\newpage
\pagebreak 
\appendix

\section{Rydberg Hamiltonian}
In the analog mode, Bloqade simulates the time evolution of a quantum state under the Schrödinger equation where the Hamiltonian is the interacting Rydberg Hamiltonian \(H\)
\begin{equation}
  i \hbar \frac{\partial}{\partial t} | \psi \rangle = H(t) | \psi \rangle  
\end{equation}
with the Hamiltonian given by:

\begin{equation}\begin{split} \label{eq:bloqade} 
    \frac{{H}(t)}{\hbar} &=  \sum_j \frac{\Omega_j(t)}{2} \left( e^{i \phi_j(t)} | 0_j \rangle \langle 1_j | + e^{-i \phi_j(t)} | 1_j \rangle \langle 0_j | \right) \\ & -\sum_j \Delta_j(t) \hat{n}_j + \sum_{j < k} V_{jk} \hat{n}_j \hat{n}_k
\end{split}\end{equation}

where \(\Omega_j(t)\) is the Rabi frequency, \(\phi_j(t)\) is the laser phase, \(\Delta_j(t)\) is the detuning, \(\hat{n}_j = |1_j\rangle \langle 1_j|\) is the number operator and \(V_{jk}\) is the interaction term between atoms \(j\) and \(k\).

Rewriting Eq. \ref{eq:bloqade} in terms of the spin operators, we have

\begin{equation}\begin{split}
    \frac{{H}(t)}{\hbar} & = \sum_j\Omega_j(t)\left[ \sigma_j^x cos(\phi_j(t))- \sigma_j^y sin(\phi_j(t))\right] 
    \\ & - \sum_j \frac{\Delta_j(t)}{2} - \sum_j \frac{\Delta_j(t)}{2} \sigma_j^z 
    \\ & + \sum_{j < k} V_{jk} \frac{(1+\sigma_j^z)}{2} \frac{(1+\sigma_k^z)}{2}
\end{split}\end{equation}

\section{Scaling Analysis}
\label{app:scaling}

In this section, we present an additional scaling analysis of our quantum reservoir computing (QRC) approach. Specifically, we investigate how the performance varies as we increase the number of qubits $N$ from 5 up to 9. We retain the same Hamiltonian structure described in the main text and measure the observables in the set $\vec{m}^{A}$, whose dimension grows with $N$. 

We consider two choices for the reservoir vector length $l_{r}$, by using the operator $\hat B^{l_r}$: one equal to the total number of elements in $\vec{m}^{A}$ (i.e., $\hat B^{l_r}=I$ ), and another set to twice that value ($l_{r} = 2 \times \dim(\vec{m}^{A})$). For each $N$, we repeat the training and testing procedures used in the main text, with the same number of training steps and random realizations of couplings.

\begin{figure}[h!btp]
    \centering
    \includegraphics[width=0.9\linewidth]{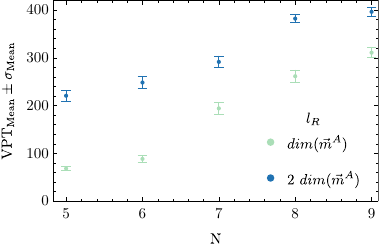}
    \caption{
    \small{VPT for the Mackey--Glass series prediction as a function of the number of qubits \(N \) for different reservoir dimension choices \(l_R\). For each point, the mean value and standard deviation of the mean are shown, based on $30$ different realizations of the Hamiltonian.
}}
    \label{fig:scaling}
\end{figure}

The results for different $N$ and $l_{r}$ values can be found in Fig.~\ref{fig:scaling}, where we plot the VPT for the Mackey--Glass prediction task for each configuration. Our results confirm that the improvements observed with $N=5$ are not restricted to a single small system. As $N$ grows, the higher-dimensional Hilbert space yields a richer set of measurements in $\vec{m}^{A}$, and the QRC performance improves for both choices of $l_{r}$. This suggests that our method can be extended to larger quantum systems with an increase of predictive power. We note that further increases in $N$ are limited mainly by computational cost in our simulations, rather than by any apparent breakdown of the QRC scheme itself.

\section{Noise Analysis}
\label{app:noise}
In order to simulate the effect of noise on the real devices, we here investigate the effect of a simple additive noise on the measured observables. Specifically, at each time step, we substitute
\[
\langle \hat{O}_i \rangle \;\longrightarrow\; \langle \hat{O}_i \rangle + \delta_i,
\]
where $\delta_i$ is sampled from a uniform distribution in the interval $[-\eta/2, \eta/2]$. This uniformly distributed noise mimics generic measurement fluctuations that could arise in actual experimental implementations.

\begin{figure}[h!btp]
    \centering
    \includegraphics[width=0.9\linewidth]{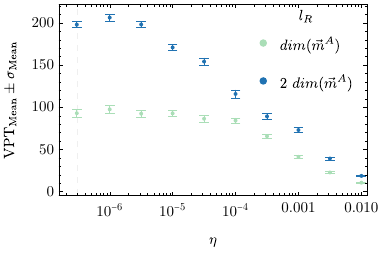}
    \caption{
    \small{VPT for the MG series prediction as a function of the noise strength \(\eta \) for two different reservoir dimension choices \(l_R\). For each point, the mean value and standard deviation of the mean are shown, based on $200$ different realizations of the Hamiltonian and random noise. The vertical dashed line indicates zero-noise simulation.
}}
    \label{fig:noise}
\end{figure}

In Fig.~\ref{fig:noise}, we show how the Valid Prediction Time (VPT) changes as a function of the noise strength~$\eta$. We report two different reservoir dimensions $l_{r}$: without interleaving zeros and doubling the length of the reservoir.  We observe that a small noise is not detrimental and could even slightly improve the performaces (blue points of  Fig.~\ref{fig:noise}).  For higher noise strenght, VPT decreases.  Overall, our results suggest that the hybrid QRC scheme retains meaningful predictive capability at small noise levels, although future investigations of more realistic noise models (such as shot noise or decoherence effects) remain an interesting avenue for further study.

\section{Comparison with a Classical Echo State Network}
\label{app:esn}
\begin{figure}[h!btp]
    \centering
    \includegraphics[width=0.9\linewidth]{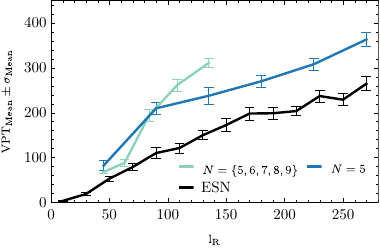}
    \caption{
    \small{VPT for the MG series prediction as a function of the trainable dimension both for classical and quantum algorithms. For each point, the mean value and standard deviation of the mean are shown, based on $30$ different realizations.
}}
    \label{fig:comparison}
\end{figure}

To benchmark our hybrid quantum-classical reservoir computing (QRC) approach against a purely classical method, we also implemented an Echo State Network (ESN) using the open-source Python library \texttt{echoes}\cite{lib_echoes}. We applied the ESN to the Mackey-Glass prediction task under the same training and testing conditions described in the main text. For a fair comparison, the ESN performances are compared by matching the dimensionality of the fitted weight vector in classical and quantum models. Consequently, in Fig.~\ref{fig:comparison}, an ESN of dimension $N_{ESN}$ is compared with a QRC whose reservoir vector also has $l_R=N_{ESN}$ components after any classical post-processing (e.g., applying $\hat{B}^{l_R}$). More specifically, the blue line is the same as the blue line in Fig.~\ref{fig:MG2} (top), and reports
the VPT value as a function of the reservoir dimension, with the number of spins $N=5$. The green line, on the other hand, reports the VPT with 5 different values of $N$ (5,6,7,8, and 9) with the reservoir dimension being equal to the total number of considered observables, i.e., 
$dim(\vec{m}^{A})$.
We see that, while the ESN’s VPT improves roughly linearly with the number of nodes, it remains consistently below the hybrid QRC approach with the same dimensionality. This finding indicates that even small-sized quantum reservoirs, combined with our classical memory augmentation, can match or outperform a standard classical ESN of similar complexity on the Mackey--Glass prediction task. Moreover, we see that the performance is better when the reservoir size is increased by scaling the number of spins (green line), rather than using post-processing (blue line).


\bibliography{QRC.bib}

\end{document}